\documentstyle[12pt]{article}
\newcommand{\be}{\begin{eqnarray}}
\newcommand{\ee}{\end{eqnarray}}
\newcommand{\ba}{\begin{array}}
\newcommand{\ea}{\end{array}}
\newcommand{\half}{{\textstyle{\frac{1}{2}}}}

\newcommand{\fourint}[1]{\int\!\frac{d^4 #1}{(2\pi)^4}}
\newcommand{\Fdual}{\widetilde{F}}

\newcommand{\partialslash}{\partial\hspace{-.5em}/\hspace{.15em}}

\newcommand{\llangle}{\left\langle}
\newcommand{\rrangle}{\right\rangle}
\newcommand{\intpsi}{\int {\cal D}\bar\psi {\cal D}\psi\,}

\begin{document}
%
%
\rightline{RUB-TPII-6/96}
\rightline{March 1996}
\rightline{hep-ph/9604324}
\vspace{.3cm}
\begin{center}
\begin{large}
{\bf Nucleon spin structure from the instanton vacuum$^\dagger$} \\
\end{large}
\vspace{1.4cm}
{\bf C. Weiss}$^{\rm 1}$ \\
\vspace{0.2cm}
{\em Institut f\"ur Theoretische Physik II \\
Ruhr--Universit\"at Bochum \\
D--44780 Bochum, Germany}
\end{center}
\vspace{1cm}
\begin{abstract}
\noindent
We discuss the evaluation of the nucleon isoscalar axial coupling, 
$g_A^{(0)}$, in the instanton vacuum, using the $1/N_c$ expansion.
This approach allows a fully consistent treatment of the $U(1)_A$ 
anomaly. We compute the nucleon matrix element of the topological 
charge density, $\langle N | F\Fdual | N \rangle$, and show that it 
reduces to the matrix element of the divergence of the isoscalar 
axial quark current. Our arguments show that the usual evaluation 
of $g_A^{(0)}$ in the chiral quark soliton model is consistent with 
the $U(1)_A$ anomaly in leading order of $1/N_c$. Such calculations
give $g_A^{(0)} = 0.36$, which is in agreement with the recent 
estimate by Ellis and Karliner.
\end{abstract}
\vspace{1cm}
\vfill
\rule{5cm}{.15mm}
\\
\noindent
{\footnotesize $\dagger$ Proceedings of the Cracow Epiphany 
Conference on Proton Structure, Cracow, Jan.\ 5--6, 1996} \\
{\footnotesize $^{\rm 1}$ E-mail: weiss@hadron.tp2.ruhr-uni-bochum.de}
%
%
%
%
\newpage
The so--called ``proton spin crisis'' has attracted a lot of interest in 
recent years \cite{EK95}. The moments of the measured polarized proton 
structure function, $g_1$, combined with information from neutron 
and hyperon beta decay, allows to extract the value of the isosinglet 
axial coupling constant of the nucleon, $g_A^{(0)}$. This quantity is
defined as the nucleon matrix element of the isosinglet axial 
current (we assume $N_f$ quark flavors),
\be
J_{5\mu}(x) &=& \sum_f^{N_f} \bar\psi_f (x) \gamma_\mu \gamma_5 
\psi_f (x) , 
\label{J_5} \\
\langle N | J_{5\mu} (0) | N \rangle &=& 
g_A^{(0)} \bar u \gamma_\mu \gamma_5 u .
\label{g_A}
\ee
In the naive parton model, $g_A^{(0)}$ can be interpreted as the
fraction of the nucleon spin carried by the quarks, commonly
denoted by $\Delta\Sigma$. The value originally obtained by EMC
was consistent with zero \cite{Ashman88}, while 
an analysis of newly available data by Ellis and Karliner comes to a value 
of $g_A^{(0)} = 0.27 \pm 0.04 \pm \ldots$ \cite{EK95}. These results have 
prompted many theoretical investigations, which provided new insights 
about the relation of the parton model language to QCD.
\par
The question about a parton interpretation aside, it remains a 
challenge to understand the small value of $g_A^{(0)}$ in QCD.
Following the work of Brodsky {\em et al.} \cite{BEK88}, who showed 
that $g_A^{(0)} = 0$ in the Skyrme model, this quantity has
been estimated in effective chiral models of the nucleon
with explicit quark degees of freedom, such as the chiral bag 
model \cite{HM88} and the chiral quark soliton 
model \cite{Wakamatsu89,BlotzPG93}, which give non-zero results
in subleading order of $1/N_c$. In these approaches one faces the 
question how the $U(1)$ axial current operator of QCD, eq.(\ref{J_5}), 
is to be represented
consistently in terms of the degrees of freedom of the effective model.
This problem does not arise for conserved currents related to 
``good'' symmetries of the strong interactions, like the electromagnetic 
or isovector axial current. The $U(1)$ axial current 
of QCD, however, has an anomalous divergence, 
\be
\partial_\mu J_{5\mu} (x) &=&
\frac{N_f}{16\pi^2} F\Fdual (x) \; 
+ \; 2 i \sum_f^{N_f} m_f \bar\psi_f \gamma_5 \psi_f (x) ,
\label{anomaly}
\ee
which is non-zero in the chiral limit, $m_f \rightarrow 0$.
Obviously, a consistent definition of the $U(1)$ axial current 
in an effective description of QCD requires an understanding of the role 
of gluonic degrees of freedom --- in particular, of topological 
fluctuations.
\par
In this note we want to show how a fully consistent and 
quantitative description of the nucleon isoscalar 
axial coupling can be achieved in the framework of the instanton 
vacuum. The relevance of instantons has been amply demonstrated
by phenomenology as well as by lattice calculations 
\cite{Sh82,varenna}. In particular, instantons describe the spontaneous 
breaking of chiral symmetry, which is the most important non-perturbative 
phenomenon determining the structure of light hadrons, including the 
nucleon \cite{varenna,DP86}. Using a $1/N_c$--expansion, 
one derives from the instanton vacuum an effective 
quark action in the form of a Nambu--Jona-Lasinio model, which
provides a convenient tool for computing correlation functions
of meson and baryon currents. This leads to a description of 
baryons as chiral solitons --- $N_c$ ``valence'' quarks moving in 
a self--consistent meson field \cite{DPP88}. Such an approach
gives a very good description of the static properties of non--strange 
and strange baryons as well as their various form factors
\cite{Goeke96}. 
\par
On the other hand, the instanton vacuum provides a microscopic
picture of the non-perturbative fluctuations of the gluon
field. It therefore offers the possibility to
include the effects of the $U(1)_A$ anomaly in an effective
description of hadrons. In the meson sector, the solution of the 
so--called $U(1)_A$ problem --- the large mass difference between 
the $\eta$ and $\eta'$ meson --- has been one of the first successes 
of the instanton vacuum \cite{tH76}. Recently, a method has been 
developed to evaluate nucleon matrix elements of gluon operators 
in the instanton vacuum. This method is based on the variational 
description of the instanton ensemble by 
Diakonov and Petrov \cite{DP84_1} and the $1/N_c$--expansion
in the quark sector. Within this approach, we can evaluate 
$g_A^{(0)}$ in a way which is fully consistent with the $U(1)_A$ anomaly.
The consistency may be demonstrated as follows. In QCD, the anomaly 
equation, eq.(\ref{anomaly}) is an operator equation, and $g_A^{(0)}$ 
may equivalently be expressed as the nucleon matrix element of the 
topological charge density, $F\Fdual (x)$,
\be
\langle N | F\Fdual (0) | N \rangle &=& 
\frac{32\pi^2}{N_f} g_A^{(0)} \, m_N \, \bar u \, i \gamma_5 u .
\label{FFdual_nucleon}
\ee
Using the method of \cite{DPW95}, we can now compute the nucleon 
matrix element of $F\Fdual (x)$ in the instanton vacuum and show 
explicitly that the value of $g_A^{(0)}$ obtained from 
eq.(\ref{FFdual_nucleon}) agrees with the nucleon matrix element of the
usual isosinglet axial quark current. In other words, the instanton vacuum
allows to evaluate the nucleon matrix element of both the left-- and the 
right--hand side of the anomaly equation, eq.(\ref{anomaly}), 
and we verify that both agree within our scheme of approximations.
\par
In the following, we give a brief outline of the approach
developed in \cite{DPW95}. We first discuss how the $U(1)_A$--anomaly
determines the dispersion of topological fluctuations of the numbers of 
instantons and antiinsantons in the ensemble.
We then formulate the prescription to calculate 
the nucleon matrix element of $F\Fdual$, and verify that the
$U(1)_A$ anomaly is realized at the level of nucleon matrix elements.
Finally, we consider the implications of this result for calculations
of $g_A^{(0)}$ within the chiral soliton 
model \cite{Wakamatsu89,BlotzPG93}.
\par
{\em $U(1)_A$--anomaly and fluctuations of the numbers of instantons.}
In the instanton approach the partition function of Euclidean 
Yang--Mills theory is reduced to a statistical ensemble of instantons 
($I$) and antiinstantons ($\bar I$). An essential property of this 
description is that the number of ``pseudoparticles'' is not fixed 
(grand canonical ensemble). This is a necessary consequence of the 
trace and $U(1)_A$ anomalies of QCD. In fact, the anomalies 
unambiguously determine the dispersion of the fluctuations of the 
number of $I$'s ($N_+$) and ${\bar I}$'s ($N_-$) in the 
ensemble. The dispersion of fluctuations of the total nuber of 
instantons, $N_+ + N_-$, is governed by the trace anomaly (see 
\cite{DPW95} for a discussion). Fluctuations of the
difference of the number of $I$'s and ${\bar I}$'s, 
\be
\Delta &\equiv& N_+ - N_- \;\; \neq \;\; 0 ,
\ee
are determined by the topological susceptibility of the vacuum,
\be
\langle \Delta^2 \rangle &=& 
\langle Q_t^2\rangle , \hspace{2cm}
Q_t \;\; \equiv \;\; \frac{1}{32\pi^2}
\int d^4 x \, F\Fdual (x) .
\label{QQ}
\ee
This follows from the fact that a single $I (\bar I)$ has topological 
charge $\pm 1$,
\be
\frac{1}{32\pi^2}
\left( \int d^4 x \, F\Fdual (x) \right)_{{\rm single}\; I (\bar I)} 
&=& \pm 1 ,
\ee
if we assume that the VEV of $Q_t^2$ is entirely due to instanton 
fluctuations.
\par
In QCD (with fermions), the topological susceptibility
is in the chiral limit determined completely by the $U(1)_A$ anomaly. 
It is possible to derive the limiting behaviour by saturating the Ward 
identity for the correlation function of two topological charge densities
with pseudoscalar meson states \cite{V79}. One 
finds ($V$ is the four--dimensional volume)
\be
\frac{\langle Q_t^2 \rangle}{V} &=& - \langle \bar\psi \psi \rangle 
\left(\sum_f^{N_f} m_f^{-1}\right)^{-1} , \hspace{2cm}
\langle \bar\psi \psi \rangle \;\; \equiv \;\; \frac{1}{N_f}\sum_f^{N_f}
\langle \bar\psi_f \psi_f \rangle.
\label{QQ_unquenched}
\ee
The topological susceptibility is proportional to the harmonic average 
of the quark masses and thus vanishes if one or more of the fermion 
flavors becomes massless. 
\par
The instanton ensemble with fermions exhibits precisely this behavior.
The vanishing of the topological susceptibility in the chiral
limit is a consequence of the presence of ``unbalanced'' zero modes 
in the fermion determinant for $N_+ \neq N_-$. In fact, a calculation
of the fermion determinant in zero--mode approximation 
(keeping only the zero mode in the interaction of the fermions 
with instantons \cite{DP86}) results in a probability distribution of 
$\Delta$ with a dispersion identical to 
eq.(\ref{QQ_unquenched}) \cite{DPW95}. 
Thus, the instanton vacuum correctly describes the topological
susceptibility of the QCD vacuum in the chiral limit.
\par
{\em Nucleon matrix element of $F\Fdual$.}
The instanton vacuum gives a quantitative prescription for
evaluating the nucleon matrix elements not only of quark,
but also of gluon operators. Specifically, we are interested in the
zero--momentum transfer nucleon matrix element of the topological
charge density, $F\Fdual (x)$. It is extracted from the correlation 
function
\be
\llangle J_N \,  J_N^\dagger \,  Q_t \rrangle ,
\label{JJQ}
\ee
where $J_N \equiv J_N (x_1 ), J_N^\dagger \equiv J_N^\dagger (x_2 )$ 
are nucleon currents consisting of $N_c$ quark fields
coupled to a color singlet. The correlation function is to be computed 
as an average over the grand canonical ensemble of instantons with 
fermions. The averaging consists of two steps: 
\begin{enumerate}
\item[{I)}] 
average over the instanton coordinates and the fermion
field of an ensemble with fixed $N_+ , N_-$ (canonical ensemble), 
allowing for $N_+ \neq N_-$.
\item[{II)}] 
average over fluctuations of $\Delta = N_+ - N_-$, with the dispersion
$\langle\Delta^2\rangle = \langle Q_t^2\rangle$
given by eq.(\ref{QQ_unquenched}).
\end{enumerate}
\par
Step I: The fixed--$N_\pm$ average can conveniently be carried out
with the help of the effective quark action, which is obtained by
integrating over the instanton coordinates 
first \cite{DPW95,DP86_prep}. It has the
form of a Nambu--Jona-Lasinio model\footnote{The fermion field
$\bar\psi$ used here is identical to $-i \psi^\dagger$ 
of \cite{DPW95}.},
\be
S_{\rm eff} [\bar\psi , \psi ] &=& \int d^4 x
\sum_f^{N_f} \bar\psi_f \partialslash \psi_f
- Y_+ - Y_-  - S_\Delta ,
\label{S_eff}
\ee
where $Y_\pm$ are $2 N_f$--fermion vertices of the form of the
't Hooft interaction (determinant in flavor),
\be
Y_\pm [\bar\psi , \psi ]
&=& \left(\frac{2V}{N}\right)^{N_f - 1} (i M)^{N_f} 
\int d^4 x \, \det_{fg} J_\pm (x)_{fg} ,
\label{Y} \\
J_\pm (x)_{fg} &=& -i \fourint{k}\fourint{l} \exp ( -i(k - l)\!\cdot\! x)
\, F(k) F(l) \, \bar\psi_f (k) \half (1 \pm \gamma_5 ) \psi_g (l) .
\nonumber 
\ee
Eq.(\ref{S_eff}) is derived by evaluating the fermion determinant
of the ensemble of $N_\pm$ instantons in saddle--point 
approximation ($1/N_c$--expansion). Here, $M$ is the dynamical quark 
mass, and $F(k)$ a form factor proportional to the wave function of the 
instanton zero mode. For unequal numbers of $I$'s and 
$\bar I$'s, the effective quark action contains a $CP$--violating 
violating fermion vertex proportional to $\Delta$,
\be
S_\Delta [\bar\psi , \psi ]
&=& - \frac{\Delta}{V \langle \bar\psi \psi \rangle} 
\left( \sum_f^{N_f} m_f^{-1} \right) (Y_+ - Y_- ) 
\;\; = \;\;
\frac{\Delta}{\langle \Delta^2 \rangle} (Y_+ - Y_- ) .
\label{delta}
\ee
This coupling of the quarks to topological fluctuations
of the number of instantons, which emerges naturally in
a saddle--point evaluation of the fermion determinant for 
$N_+ \neq N_-$, plays a crucial role in the realization of 
the $U(1)_A$ anomaly, as seen below.
An effective action similar to eqs.(\ref{S_eff}, \ref{delta}) 
has also been suggested by Nowak {\em et al.} \cite{Nowak89}. 
These authors introduce a local (space--time dependent) 
density of instantons through a coarse--graining procedure. 
We do not require this concept in our approach. 
\par
With the help of the effective quark action, eq.(\ref{S_eff}), the 
fixed--$N_\pm$ average of two nucleon currents can be 
represented as
\be
\llangle J_N \, J_N^\dagger \rrangle_{{\rm fixed}-N_\pm}
&=& 
\frac{{\displaystyle \intpsi J_N \, J_N^\dagger \,
\exp \left( -S_{\rm eff}[\bar\psi , \psi ] \right) }}
{{\displaystyle \intpsi
\exp \left( -S_{\rm eff}[\bar\psi , \psi ] \right) }} 
\;\; \equiv \;\;
\llangle J_N \, J_N^\dagger \rrangle_{{\rm eff}} .
\ee
In the limit of large Euclidean times this correlation function
is dominated by the saddle point corresponding to the
static chiral soliton \cite{DPP88}. Also correlation functions
with gluon operators can systematically be expressed as averages 
in the effective quark theory \cite{DPW95}. 
In the case of the topological charge,
eq.(\ref{JJQ}) the result is particularly simple,
\be
\llangle J_N \, J_N^\dagger \, Q_t
\rrangle_{{\rm fixed}-N_\pm}
&=& \Delta \llangle J_N \, J_N^\dagger \, \rrangle_{{\rm eff}} .
\label{NJJ}
\ee
The free nucleon correlation function is simply multiplied 
by the topological charge of the fixed--$N_\pm$ instanton ensemble, 
$\Delta$, which seems intuitively plausible.
\par
Step II: Having computed the average over the fixed--$N_\pm$ ensemble, we 
now must average the result over fluctuations of the topological charge,
$\Delta$,
\be
\llangle J_N \, J_N^\dagger \, Q_t
\rrangle
&\equiv& \sum_\Delta P(\Delta ) 
\llangle J_N \, J_N^\dagger \, Q_t
\rrangle_{{\rm fixed}-N_\pm} ,
\label{grand}
\ee
where $P(\Delta )$ is the probability distribution of $\Delta$ 
in the grand ensemble, describing the dispersion
eq.(\ref{QQ_unquenched}). Since the dispersion is $O(m)$ and 
thus small in the chiral limit, it is sufficient to evaluate the 
fixed--$N_\pm$ average on the RHS of eq.(\ref{grand}) to 
the first non--vanishing order in $\Delta$. For our
correlation function eq.(\ref{NJJ}) this means
\be
\llangle J_N \, J_N^\dagger \, Q_t
\rrangle
&=& \langle \Delta^2 \rangle \frac{d}{d \Delta}
\llangle J_N \, J_N^\dagger 
\rrangle_{{\rm eff}} \,
\rule[-.3cm]{.15mm}{.7cm} \; 
\rule[-.2cm]{0mm}{0cm}_{\Delta = 0} 
\label{grand_1} \\
&=& \llangle J_N \, J_N^\dagger \, (Y_+ - Y_- )
\rrangle_{{\rm eff}} \,
\rule[-.3cm]{.15mm}{.7cm} \; 
\rule[-.2cm]{0mm}{0cm}_{\Delta = 0} \\
&=& \llangle J_N \, J_N^\dagger \,
\frac{1}{2 N_f}\int d^4 x\, \partial_\mu J_{5\mu} (x) 
\rrangle_{{\rm eff}} \,
\rule[-.3cm]{.15mm}{.7cm} \; 
\rule[-.2cm]{0mm}{0cm}_{\Delta = 0} .
\label{final}
\ee
Note that the factor $\langle \Delta^2 \rangle$ has canceled with a 
corresponding factor in eq.(\ref{delta}), so that the result is
finite in the chiral limit. In eq.(\ref{final}) we 
have used the fact the $Y_+ - Y_-$ is nothing but 
the divergence of the isosinglet axial current of the effective
quark theory, eq.(\ref{S_eff}),
\be
\int d^4 x\, \partial_\mu J_{5\mu}(x) &=& 2 N_f (Y_+ - Y_- ) ,
\ee
as follows from the equations of motion in leading order of $1/N_c$.
\par
Eq.(\ref{final}) now shows how the $U(1)_A$--anomaly is realized:
the grand canonical correlation function with the topological
charge reduces to the Nambu--Jona-Lasinio correlation function with an
insertion of the divergence of the usual isosinglet axial current
of the massive ``constituent'' quark field. 
Taking the limit of large Euclidean times, one obtains from 
eq.(\ref{final}) the corresponding equation for the nucleon matrix 
elements.
\par
{\em Conclusions.} We have demonstrated how the instanton vacuum 
naturally includes the $U(1)_A$--anomaly in the effective description 
of the nucleon. The mechanism are the fluctuations of the difference 
of the number of $I$'s and $\bar I$'s in the ensemble, the dispersion of 
which is governed by the $U(1)_A$--anomaly. Our argument shows that a 
calculation of $g_A^{(0)}$ in the chiral quark soliton model may 
be regarded as a consistent estimate of the nucleon matrix 
element of the topological charge, eq.(\ref{FFdual_nucleon}), in the 
instanton vacuum. We have established this equality in the
leading order of the $1/N_c$ expansion. A systematic treatment 
of higher--order corrections in $1/N_c$ is in principle possible, 
but difficult. For instance, the 't Hooft form of the effective
quark interaction, eq.(\ref{Y}), applies only to the leading 
order in $1/N_c$.
\par
Calculations in the chiral quark soliton model with $SU(3)$ flavor
give $g_A^{(0)} \sim 0.36$ \cite{BlotzPG93}. This result is in 
agreement with the value obtained in the recent analysis by Ellis and 
Karliner \cite{EK95}.
\par
The correct description of the $U(1)_A$ anomaly is a crucial test 
for the general method for evaluating matrix elements of gluon 
operators developed in \cite{DPW95}. This method allows also
to evaluate the nucleon matrix elements of QCD operators of
leading and non-leading twist, as are encountered in the OPE--description
of deep--inelastic scattering. Such an approach opens the prospect
of describing the deep--inelastic structure of the nucleon
consistently with its hadronic structure, in one well-defined
scheme of approximations. Calculations of
the nucleon structure functions using this method are in progress.
\vspace{1cm}
\par
The ideas presented here have been developed together with
D.I.\ Diakonov and M.V.\ Polyakov. It is my pleasure to thank
them for a stimulating collaboration. I am also indebted to 
P.V.\ Pobylitsa for many interesting discussions.
%
%

\end{document}